\documentclass{icrc2009}

\usepackage{graphicx}   
\usepackage[caption=false,font=footnotesize]{subfig}
\usepackage{fixltx2e}
\usepackage{url}

\newcommand{\shorttitle}[1]%
{\markboth{Proceedings of the 31\MakeLowercase{$^{st}$} ICRC, {\L}\'{o}d\'{z} 2009}{#1} }

\begin{document}
\title{Atmospheric Variations as observed by IceCube}

\author{
\IEEEauthorblockN{
Serap Tilav\IEEEauthorrefmark{1},
Paolo Desiati\IEEEauthorrefmark{2},
Takao Kuwabara\IEEEauthorrefmark{1},
Dominick Rocco\IEEEauthorrefmark{2},\\
Florian Rothmaier\IEEEauthorrefmark{3},
Matt Simmons\IEEEauthorrefmark{1},
Henrike Wissing\IEEEauthorrefmark{4}\IEEEauthorrefmark{5} for the IceCube Collaboration\IEEEauthorrefmark{6}
}\\
\IEEEauthorblockA{\IEEEauthorrefmark{1}
 Bartol Research Institute and Dept. of Physics and Astronomy, University of Delaware, Newark, DE 19716, USA.}
\IEEEauthorblockA{\IEEEauthorrefmark{2}
Dept. of Physics, University of Wisconsin, Madison, WI 53706, USA.}
\IEEEauthorblockA{\IEEEauthorrefmark{3}
Institute of Physics, University of Mainz, Staudinger Weg 7, D-55099 Mainz, Germany.}
\IEEEauthorblockA{\IEEEauthorrefmark{4}
III Physikalisches Institut, RWTH Aachen University, D-52056 Aachen, Germany.}
\IEEEauthorblockA{\IEEEauthorrefmark{5}
Dept. of Physics, University of Maryland, College Park, MD 20742, USA.}
\IEEEauthorblockA{\IEEEauthorrefmark{6}
See the special section of these proceedings}
}
\shorttitle{Atmospheric Variation}
\maketitle

\begin{abstract}
We have measured the correlation of rates in IceCube with long and short term variations in the South Pole atmosphere.
The yearly temperature variation in the middle stratosphere (30-60 hPa) is highly correlated with the high energy 
muon 
rate observed deep in the ice, and causes a $\pm$10\% seasonal modulation in the event rate.
The counting rates of the surface detectors, which are due to secondary particles of relatively low energy 
(muons, electrons and photons), have a negative correlation with temperatures in the lower layers of the 
stratosphere (40-80 hPa), and are modulated at a level of $\pm$5\%. The region of the atmosphere between pressure levels 20-120 hPa,
where the first cosmic ray interactions occur and the produced pions/kaons interact or decay to muons, is the Antarctic ozone layer. 
The anti-correlation between surface and deep ice trigger 
rates reflects the properties of pion/kaon decay and interaction as the density of the stratospheric ozone layer changes.
Therefore, IceCube closely probes the ozone hole dynamics, and the temporal behavior of the stratospheric temperatures.
\end{abstract}


\section{Introduction}
The IceCube Neutrino Observatory, located at the geographical South Pole (altitude 2835m), 
has been growing incrementally in size since 2005, surrounding its predecessor AMANDA\cite{Karle}.
As of March 2009, IceCube consists of 59 strings in the Antarctic ice, 
and 59 stations of the IceTop cosmic ray air shower array on the surface.
Each IceCube string consists of 60 Digital Optical Modules (DOMs) deployed at depths of 1450-2450m, and
each IceTop station comprises 2 ice Cherenkov tanks with 2 DOMs in each tank.

IceCube records the rate of atmospheric muons with E$_\mu \geq$ 400 GeV. Muon events pass the IceCube 
Simple Majority Trigger (InIce\_SMT) if 8 or more DOMs are triggered in 5$\mu$sec. 
The IceTop array records air showers with primary cosmic ray energies above 300 TeV. 
Air showers pass the IceTop Simple Majority Trigger (IceTop\_SMT) if 6 or more surface DOMs trigger within 5$\mu$sec. 

As the Antarctic atmosphere goes through seasonal changes, the characteristics of the cosmic ray interactions 
in the atmosphere follow these variations. When the primary cosmic ray interacts with atmospheric nuclei, 
the pions and kaons produced at the early interactions mainly determine the nature of the air shower. 
For the cosmic ray energies discussed here, these interactions happen in the ozone layer 
(20-120 hPa pressure layer at 26 down to 14 km) of the South Pole stratosphere. 
During the austral winter, when the stratosphere is cold and dense, 
the charged mesons are more likely to interact and produce secondary low energy particles. 
During the austral summer when the warm atmosphere expands and becomes less dense, 
the mesons more often decay rather than interact.

\begin{figure*}[th]
\begin{center}
\includegraphics[width=1.0\textwidth]{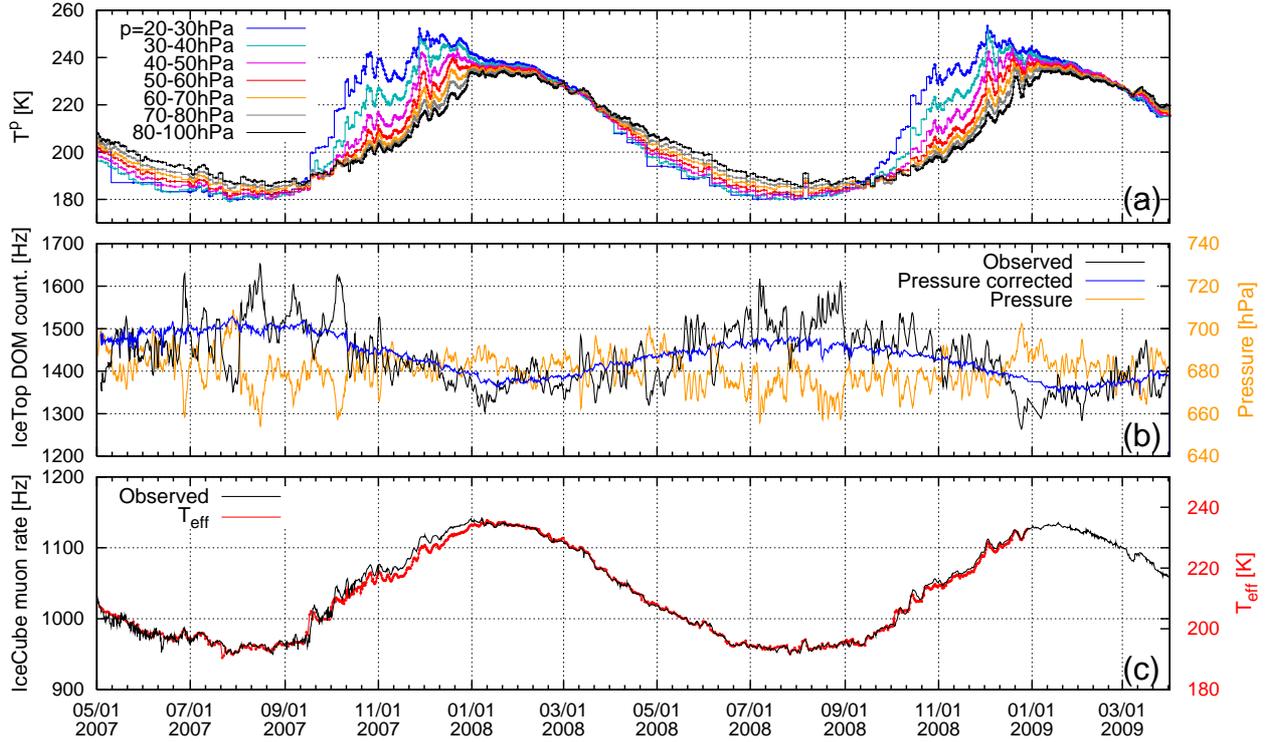}
\end{center}
\caption{
The temporal behavior of the South Pole stratosphere from May 2007 to April 2009 is compared to IceTop 
DOM counting rate and the high energy muon rate in the deep ice.
(a) The temperature profiles of the stratosphere at pressure layers from 
20 hPa to 100 hPa where the first cosmic ray interactions happen. 
(b) The IceTop DOM counting rate (black -observed, blue -after barometric correction) and the surface 
pressure (orange). (c) The IceCube muon trigger rate and the calculated effective temperature (red).
}
\end{figure*}

Figure 1 demonstrates the modulation of rates in relation to the temporal changes of the South Pole stratosphere. 
The scaler rate of a single IceTop DOM on the surface is mostly due to low energy secondary particles 
(MeV electrons and gammas, $\sim$1 GeV muons)\cite{Clem 2008}
produced throughout the atmosphere, and therefore highly modulated by the atmospheric pressure. 
However, after correcting for the barometric effect, the IceTop DOM counting rate also reflects 
the initial pion interactions in the middle stratosphere. 
The high energy muon rate in the deep ice, 
on the other hand, directly traces the decay characteristics of high energy pions in higher layers of  
the stratosphere. 
The IceCube muon rate reaches its maximum at the end of January when the stratosphere is warmest 
and most tenuous. 
Around the same time Icetop DOMs measure the lowest rates on the surface as the pion interaction 
probability reaches its minimum. The high energy muon rate starts to decline as the stratosphere 
gets colder and denser. The pions will interact more often than decay, 
yielding the maximum rate in IceTop and the minimum muon rate in deep ice at the end of July. 

\section{Atmospheric effects on the IceCube Rates}
The Antarctic atmosphere is closely monitored by the NOAA Polar Orbiting Environmental Satellites (POES)
 and by the radiosonde balloon launches of the South Pole Meteorology Office. 
However, stratospheric data is sparse during the winter when the balloons do not reach high altitudes, 
and satellite based soundings fail to return reliable data. 
For such periods NOAA derives temperatures from their models.
We utilize both the ground-based data and satellite measurements/models for our analysis.

\subsection{Barometric effect}
In first order approximation, the simple correlation 
between log of rate change $\Delta\{ln R\}$ and the surface pressure change $\Delta P$ is
\begin{eqnarray}
 \Delta\{ln R\}=\beta\cdot\Delta P
\end{eqnarray}
where $\beta$ is the barometric coefficient.

As shown by the black line in the Figure 1b, the observed IceTop DOM counting rate varies by $\pm$10\% 
in anti-correlation with surface pressure, and 
the barometric coefficient is determined to be $\beta = -0.42$\%/hPa. 
Using this value, the pressure corrected scaler rate is plotted as the smoother line (blue) in Figure 1b.
The cosmic ray shower rate detected by the IceTop array also varies by $\pm$17\% in anti-correlation with
surface pressure, and can be corrected with a $\beta$ value of $-0.77$\%/hPa.
As expected \cite{Dorman}, the IceCube muon rate shown in Figure 1c is not correlated with surface pressure. 
However, during exceptional stratospheric temperature changes, the second order temperature effect on 
pressure becomes large enough to cause anti-correlation of the high energy muon rate 
with the barometric pressure. During such events the effect directly reflects sudden 
stratospheric density changes, specifically in the ozone layer.

\subsection{Seasonal Temperature Modulation}

Figure 1 clearly demonstrates the seasonal temperature effect on the rates. 
The IceTop DOM counting rate, after barometric correction, shows $\pm$5\% negative temperature 
correlation. On the other hand, the IceCube muon rate is positively correlated with $\pm$10\% seasonal variation.

From the phenomenological studies \cite{Barrett 1952}\cite{Ambrosio 1997}, 
it is known that correlation between temperature and muon intensity can be 
described by the effective temperature $T_{eff}$, defined by the weighted 
average of temperatures from the surface to the top of the atmosphere.
$T_{eff}$ approximates the atmosphere as an isothermal body, 
weighting each pressure layer according to its relevance to muon production 
in atmosphere \cite{Ambrosio 1997}\cite{Gaisser 1990}.

The variation of muon rate \mbox{$\Delta R_\mu/<R_\mu>$} is related to the effective temperature as
\begin{eqnarray}
 \frac{\Delta R_\mu}{<R_\mu>}
 =\alpha_T\frac{\Delta T_{eff}}{<T_{eff}>},
\end{eqnarray}
where $\alpha_T$ is the atmospheric temperature coefficient. 

Using balloon and satellite data for the South Pole atmosphere, we calculated the effective 
temperature as the red line in Figure 1c. We see that it traces the IceCube muon rate remarkably well.
The calculated temperature coefficient $\alpha_T=0.9$ for the IceCube muon rate agrees well with the expectations of 
models as well as with other experimental measurements\cite{Grashorn 2008}\cite{Osprey 2009}.

In this paper, we also study in detail the relation between rates and stratospheric temperatures 
for different pressure layers from 20 hPa to 100 hPa as
\begin{eqnarray}
\frac{\Delta R}{<R>}=\alpha_T^p\frac{\Delta T^p}{<T^p>}.
\end{eqnarray}

The temperature coefficients for each pressure layer $\alpha_T^p$ and the correlation coefficient 
$\gamma$ are determined from regression analysis.
Pressure-corrected IceTop DOM counting rate, and IceCube muon rate are sorted 
in bins of $\sim$ 10 days, and deviations \mbox{$\Delta R/<R>$} 
from the average values are compared with the deviation of 
temperatures at different depths \mbox{$\Delta T^p/<T^p>$}.

We list the values of $\alpha_T^p$ and $\gamma$ for IceCube muon rate and IceTop DOM counting rate in Table 1.
We find that the IceCube muon rate correlates best with the temperatures of 30-60 hPa pressure layers,
while the IceTop DOM counting rate shows the best correlation with 60-80 hPa layers.
In Figure 2. we plot the rate and temperature correlation for layers which yield the best correlation.

\begin{table}[t]
\caption{{\rm Temperature and correlation coefficients of rates for different stratospheric layers 
of 20-100 hPa and $T_{eff}$.}}
{\scriptsize
\begin{tabular}{|cc|cccc|}
\hline
 &       & &       & &       \\
      &       &\multicolumn{2}{c}{IceCube Muon} & \multicolumn{2}{c|}{IceTop Count.}\\
\hline
\hline
 &       & &       & &       \\
$P$    &$<T^p>$  & $\alpha_T^p$ & $\gamma$     & $\alpha_T^p$ & $\gamma$\\
 $(hPa)$ &$(K)$     &              &              &               & \\ 
\hline
20-30 & 214.0 & 0.512 & 0.953 & -0.194 & -0.834\\
30-40 & 208.7 & 0.550 & 0.986 & -0.216 & -0.906\\
40-50 & 207.3 & 0.591 & 0.993 & -0.240 & -0.946\\
50-60 & 206.6 & 0.627 & 0.985 & -0.261 & -0.968\\
60-70 & 206.3 & 0.656 & 0.971 & -0.278 & -0.975\\
70-80 & 206.3 & 0.679 & 0.954 & -0.292 & -0.975\\
80-100& 206.5 & 0.708 & 0.927 & -0.310 & -0.971\\
\hline
\hline
 &       & &       & &       \\
   &$<T_{eff}>$ & $\alpha_T$  & $\gamma$     & $\alpha_T$  & $\gamma$\\
\hline
&       & &       & &      \\
      & 211.3 & 0.901 & 0.990 & -0.360 & -0.969\\
\hline
\end{tabular}
}
\end{table}
\begin{figure}[t]
\begin{center}
\includegraphics[width=0.5\textwidth]{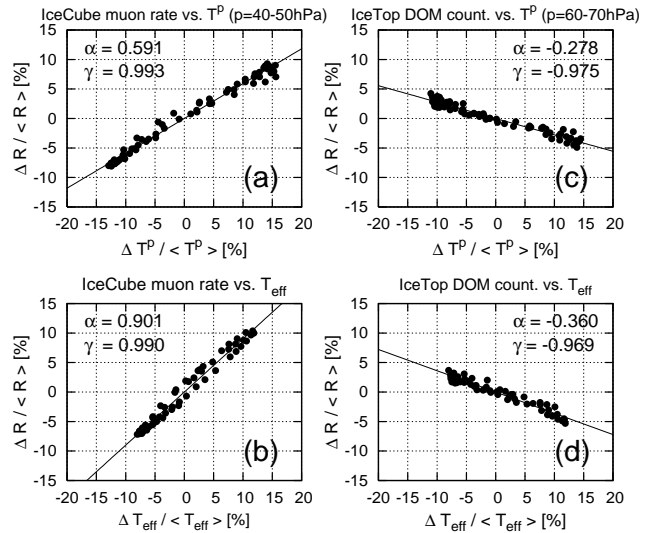}
\end{center}
\caption{Correlation of IceCube muon and IceTop DOM counting rates with stratospheric temperatures and $T_{eff}$.
(a) IceCube muon rate vs. temperature at 40-50 hPa pressure layer. 
(b) IceCube muon rate vs. effective temperature. 
(c) IceTop DOM counting rate vs. temperature at 70-80 hPa pressure layer. 
(d) IceTop DOM counting rate vs. effective temperature.}
\end{figure}

\section{Exceptional Stratospheric Events and the Muon rates}
The South Pole atmosphere is unique because of the polar vortex. 
In winter a large-scale counter clockwise flowing cyclone forms over 
the entire continent of Antarctica, isolating the Antarctic atmosphere from higher latitudes.
Stable heat loss due to radiative cooling continues until August without much disruption, 
and the powerful Antarctic vortex persists until the sunrise in September. 
As warm air rushes in, the vortex loses its strength, shrinks in size, 
and sometimes completely disappears in austral summer. The density profile inside the vortex
changes abruptly during the sudden stratospheric warming events, which eventually may cause the
vortex collapse. The ozone depleted layer at 14-21 km altitude (ozone hole),
observed in September/October period, is usually replaced with ozone rich layer at 18-30 km
soon after the vortex breaks up.  

Apart from the slow seasonal temperature variations, IceCube also probes the atmospheric density changes due 
to the polar vortex dynamics and vigorous stratospheric 
temperature changes on time scales as short as days or even hours, which are of great meteorological interest.  

An exceptional and so far unique stratospheric event has already been observed in muon data taken with 
IceCube's predecessor AMANDA-II. 

\subsection{2002 Antarctic ozone hole split detected by AMANDA}
\begin{figure}[h]
\begin{center}
\includegraphics[width=0.5\textwidth]{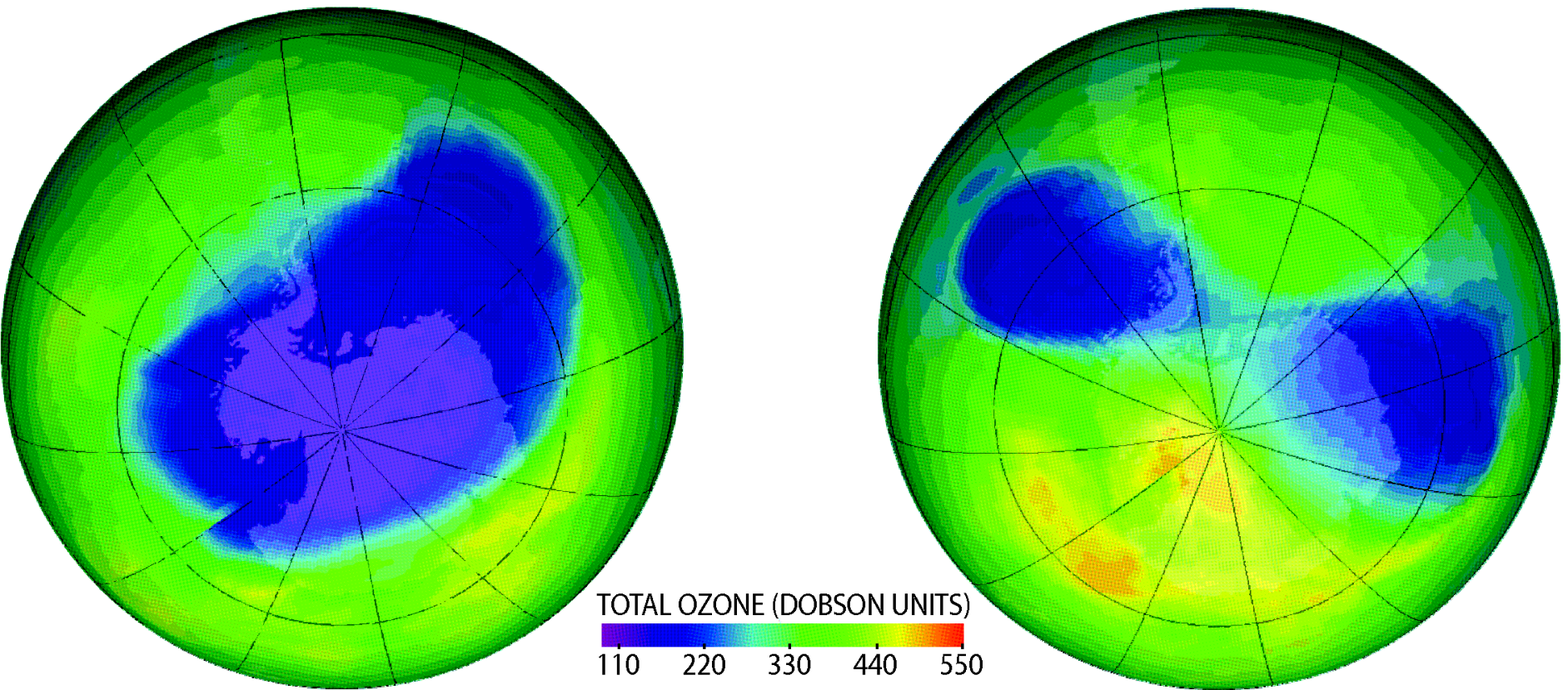}
\end{center}
\caption{Ozone concentration over the southern hemisphere on September 20th 2002 (left) 
and September 25th 2002 (right) \cite{nasaozonewatch}.}
\begin{center}
\includegraphics[width=0.4\textwidth]{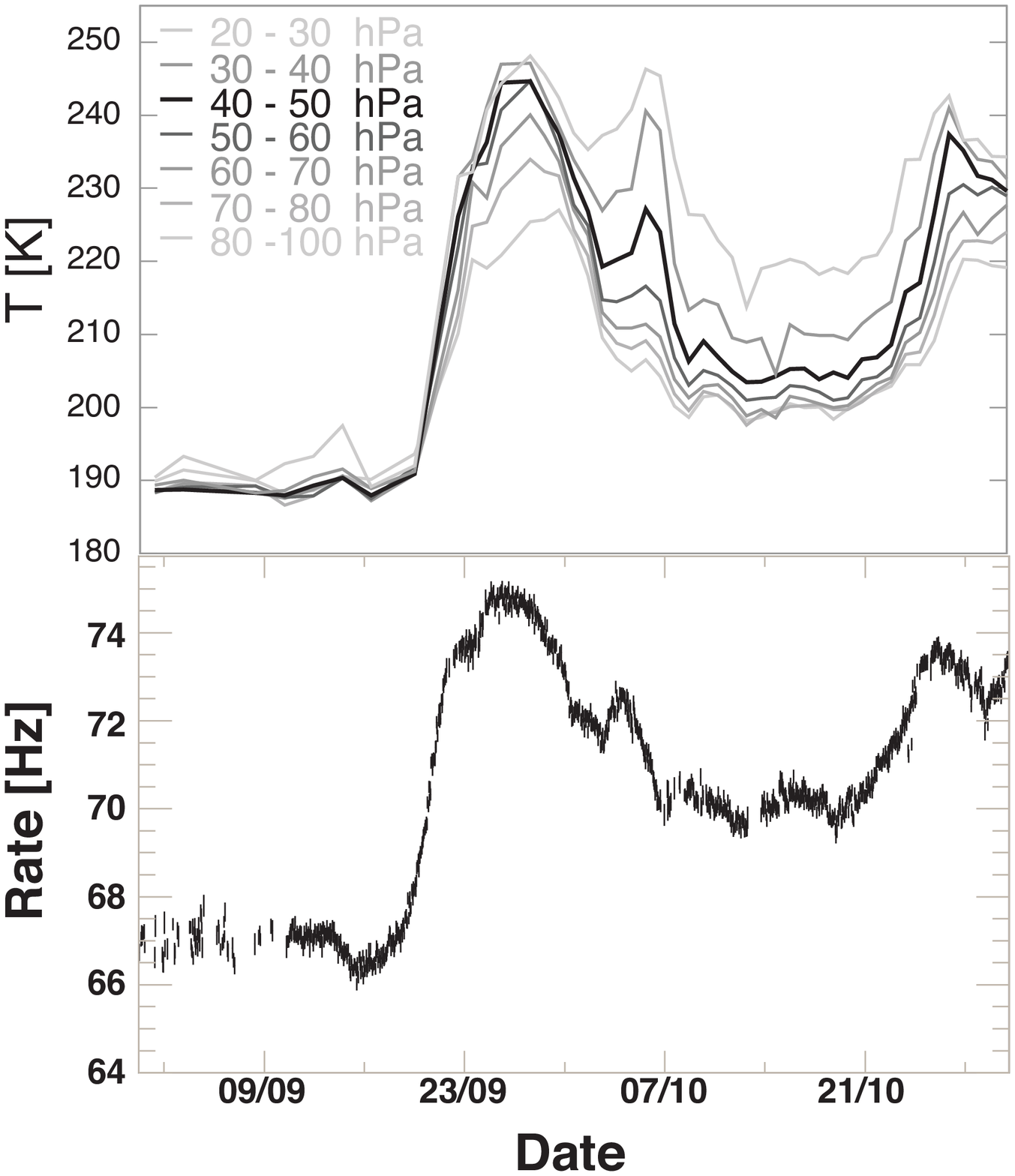}
\end{center}
\caption{Average temperatures in various atmospheric layers over the South Pole (top) 
and deep ice muon rate recorded with the AMANDA-II detector (bottom) during the Antarctic ozone hole split of September-October 2002.}
\end{figure}

In late September 2002 the Antarctic stratosphere underwent its first recorded major Sudden Stratospheric Warming (SSW), 
during which the atmospheric temperatures increased by 40 to 60 K in less than a week. This unprecedented event caused the
polar vortex and the ozone hole, normally centered above the South Pole, to split into two smaller, 
separate off-center parts (Figure 3) \cite{varotsos}. 

Figure 4 shows the stratospheric temperatures between September and October 2002 along with the AMANDA-II muon rate.
The muon rate traces temperature variations in the atmosphere in great detail, 
with the strongest correlation observed for the 40-50\,hPa layer.
\subsection{South Pole atmosphere 2007-2008}
\begin{figure}[h]
\begin{center}
\includegraphics[width=0.5\textwidth]{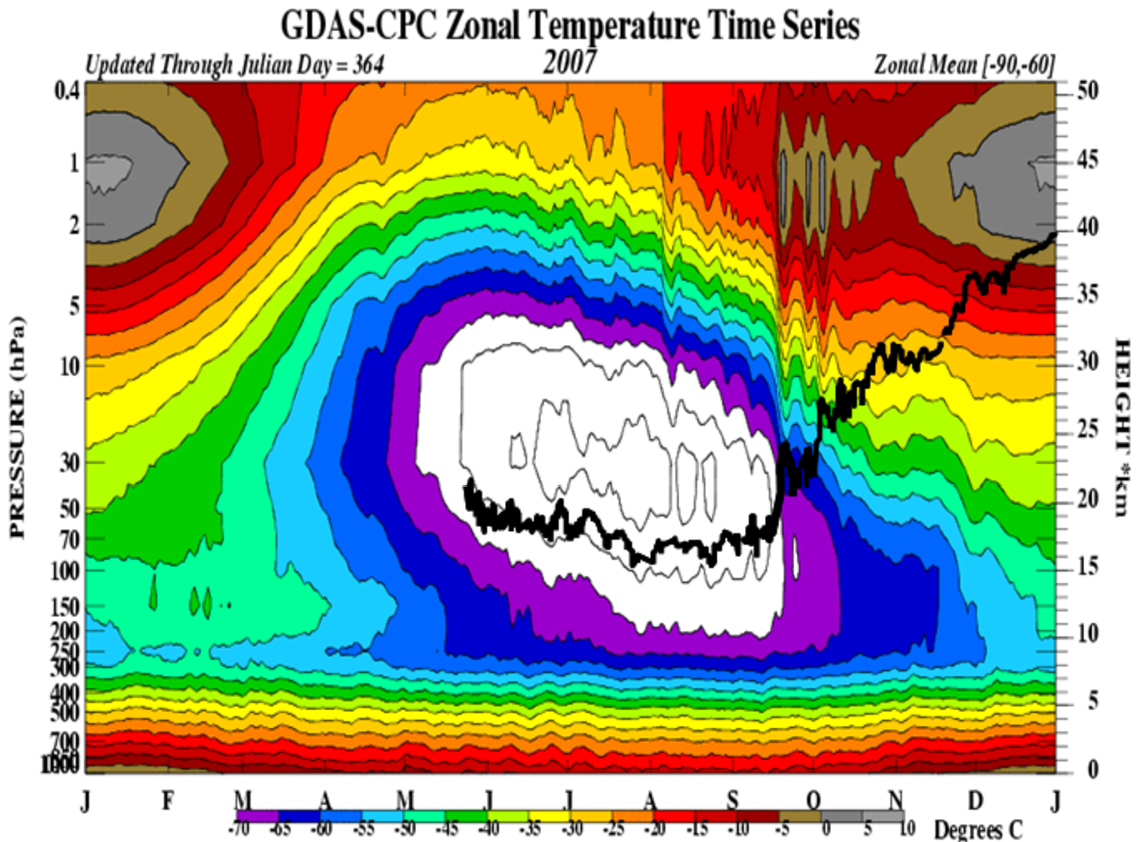}
\\\vspace{0.5cm}
\includegraphics[width=0.5\textwidth]{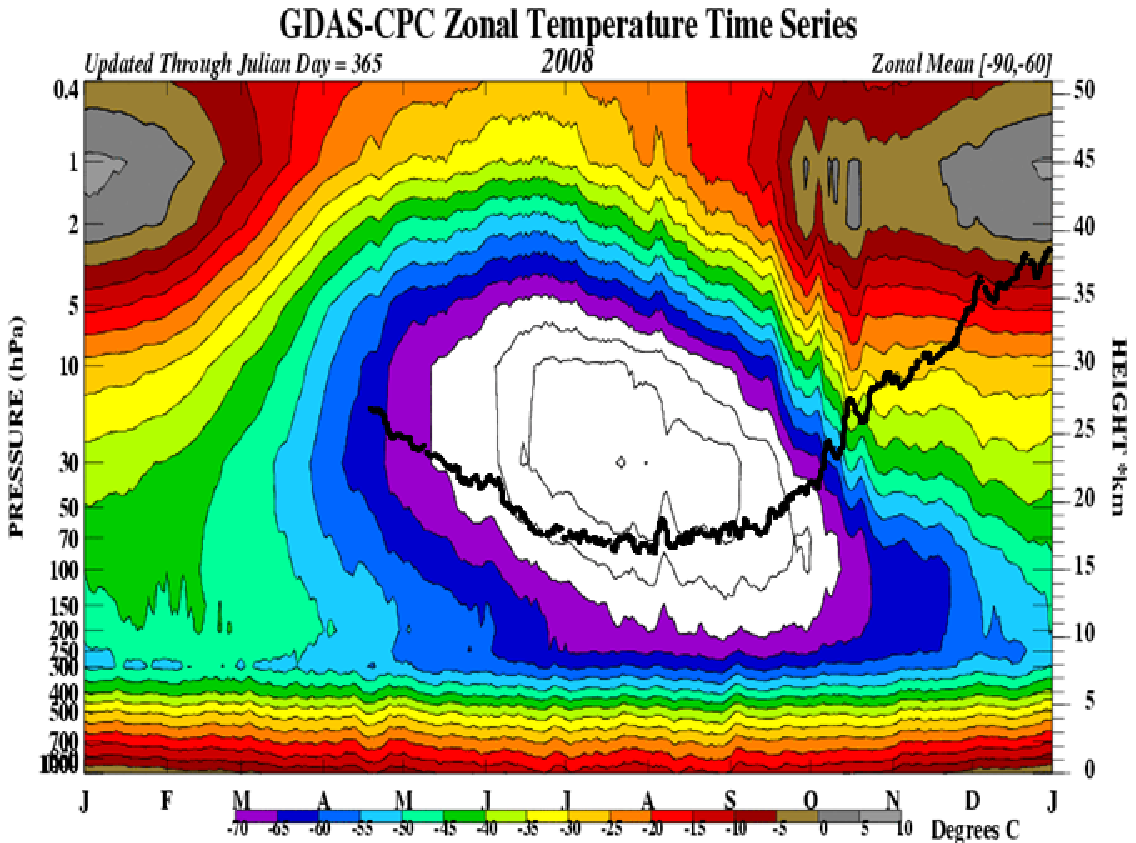}
\end{center}
\caption{The temperature time series of the Antarctic atmosphere produced by NOAA\cite{stratosphere} are 
shown for 2007 and 2008. 
The pattern observed in the deep ice muon rate (black line) is superposed onto the plot to display the
striking correlation with the stratospheric temperature anomalies.}
\end{figure}

Unlike in 2002, the stratospheric conditions over Antarctica were closer to average in 2007 and 2008. 
In 2007 the polar vortex was off-center from the South Pole during most of September and October,
resulting in greater heat flux into the vortex, which decreased rapidly in size. 
When it moved back over the colder Pole region in early November it gained strength and persisted 
until the beginning of December.
The 2008 polar vortex was one of the largest and strongest observed in the last 10 years over the South Pole. 
Because of this, the heat flux entering the area was delayed by 20 days. 
The 2008 vortex broke the record in longevity by persisting well into mid-December. 
 
In Figure 5 we overlay the IceCube muon rate over the temperature profiles of the Antarctic 
atmosphere produced by the NOAA Stratospheric analysis team\cite{stratosphere}. We note that the anomalous 
muon rates (see, for example, the sudden increase by 3\% on 6 August 2008) observed by IceCube
are in striking correlation with the middle and lower stratospheric temperature anomalies.

We are establishing automated detection methods for anomalous events in the South Pole atmosphere 
as well as a detailed understanding and better modeling of cosmic ray interactions during such
stratospheric events. 
\section{Acknowledgements}
We are grateful to the South Pole Meteorology Office and the Antarctic Meteorological Research Center of the
University of Wisconsin-Madison for providing the meteorological data.
This work is supported by the National Science Foundation.

\end{document}